\documentclass[conference]{IEEEtran}
\IEEEoverridecommandlockouts
\usepackage{cite}
\usepackage{amsmath,amssymb,amsfonts}
\usepackage{graphicx}
\graphicspath{{img/}}      
\usepackage{booktabs}
\usepackage{colortbl}
\usepackage{textcomp}
\usepackage{xcolor}
\definecolor{flagbg}{HTML}{FCE4E4}   
\definecolor{goodbg}{RGB}{226,239,218}
\usepackage[hidelinks]{hyperref}
\usepackage{url}
\usepackage{orcidlink}
\usepackage{stfloats}

\usepackage{helvet}   
\usepackage{tikz}
\usetikzlibrary{arrows.meta,positioning,calc,shadows}
\usepackage{pifont}
\usepackage[misc]{ifsym}

\begin{document}

\title{Same Speaker, Not Human: Forensic Voice Examiners and Modern Deepfake Speech
\thanks{\textsuperscript{\Letter} Corresponding author.}
\thanks{This work was supported by the Brno University of Technology internal project FIT-S-26-9011.}
}

\author{
\IEEEauthorblockN{Kamil Malinka\,\orcidlink{0000-0002-9009-2193}\,\textsuperscript{\Letter}, 
Anton Firc\,\orcidlink{0000-0002-4717-1910}, 
Milan \v{S}alko\,\orcidlink{0009-0004-9604-168X}, 
Martin Pere\v{s}\'{i}ni\,\orcidlink{0000-0002-2875-9567},
and Vojt\v{e}ch Stan\v{e}k\,\orcidlink{0009-0000-5722-0571}}
\IEEEauthorblockA{
\textit{Security@FIT, Brno University of Technology} \\
Brno, Czech Republic \\
\{malinka, ifirc, isalko, iperesini, istanek\}@fit.vut.cz}
}

\maketitle

\begin{abstract}
Synthetic speech can now imitate a person closely enough for a cloned recording to be presented as evidence that the person spoke in a criminal case. When this claim is disputed, a forensic voice examiner compares the recording with a genuine reference and reports a likelihood ratio (LR) for the same-speaker versus different-speaker propositions. Yet the LR does not test whether the speech is human or synthetic. We tested this gap with two practicing forensic voice examiners. Both completed protocol-matched English and Czech listening tasks, and Examiner~A performed a casework-style workup on the Czech set. Each recording contained four sentences and was genuine, fully synthetic, or partially synthetic with one synthetic sentence. The examiners detected 94 percent of YourTTS full-spoof sentences in English and 84 percent in Czech, compared with 65 and 63 percent for ElevenLabs v1. The similar decline across languages makes second-language difficulty alone an unlikely explanation. In the forensic workup, two of four ElevenLabs full spoofs received LRs that overlapped those of the four genuine recordings and were not flagged for synthesis in the final conclusion; three of four ElevenLabs partial spoofs were also not flagged. Synthetic speech can therefore receive same-speaker support even when synthesis is explicitly considered. Deepfake-detector scores should enter forensic opinions only when the detector has been validated and calibrated for conditions representative of the case; otherwise, recording origin should remain inconclusive.
\end{abstract}

\begin{IEEEkeywords}
audio deepfakes, forensic voice comparison, likelihood ratio, synthetic speech, partial spoofing
\end{IEEEkeywords}

\section{Introduction}\label{sec:intro}

Criminal cases may rely on voice recordings: an intercepted call, a threat left on voicemail, or a recording that places a suspect at a scene. When the speaker’s identity is disputed, a forensic voice examiner compares the questioned recording with known material and provides an expert opinion to the court. Synthetic speech threatens this process because a cloned voice can falsely implicate a person while still sounding like them. Voice cloning has already been used to authorize a fraudulent wire transfer~\cite{fraud} and has been identified as a means of planting fabricated evidence in criminal cases~\cite{heliyon}. Yet automatic detectors that report near-zero error on benchmarks often fail on unseen data~\cite{datasetaudit,lastmile,mullergen}, while ordinary listeners detect modern fakes only slightly above chance~\cite{diel,mai}. Courts must therefore be prepared for deepfakes to appear as voice evidence. This paper asks whether a forensic workflow designed to establish \emph{speaker identity} also reveals the recording’s \emph{synthetic origin}.

Forensic voice comparison normally asks a narrower question: does the questioned recording support the same-speaker or the different-speaker proposition relative to a trusted reference~\cite{enfsiguide,morrison19}? The resulting likelihood ratio~(LR) expresses the relative strength of support for those propositions~\cite{enfsi,morrison19}; it does not test whether the speech was produced by a human. \textit{A cloned voice may therefore receive strong same-speaker support while remaining synthetic.} Origin requires separate evidence.

\begin{figure}[t]
\centering
\sffamily
\definecolor{neutral}{HTML}{4C5B6B}
\definecolor{s1}{HTML}{2C6FBB}
\definecolor{good}{HTML}{1B9E77}
\definecolor{bad}{HTML}{D95F02}
\newcommand{\cmark}{\textcolor{good}{\ding{51}}}
\newcommand{\xmark}{\textcolor{bad}{\ding{55}}}
\newcommand{\pf}{\tikz[baseline=-0.4ex]{\fill[good] (0,1.5mm) circle (0.95mm);
  \fill[good,rounded corners=0.5mm] (-1.4mm,-2mm) rectangle (1.4mm,0.4mm);}}
\newcommand{\po}{\tikz[baseline=-0.4ex]{\draw[neutral!50,line width=0.5pt] (0,1.5mm) circle (0.95mm);
  \draw[neutral!50,line width=0.5pt,rounded corners=0.5mm] (-1.4mm,-2mm) rectangle (1.4mm,0.4mm);}}
\begin{tikzpicture}[
  font=\scriptsize,
  bx/.style={rounded corners=3pt,draw=#1,line width=0.8pt,fill=#1!6,align=center,inner sep=3.5pt,
             drop shadow={shadow xshift=0.4pt,shadow yshift=-0.5pt,opacity=0.16,fill=neutral}},
  bx/.default=neutral,
  ctag/.style={fill=neutral,text=white,font=\bfseries\scriptsize,inner sep=1.4pt,rounded corners=1.5pt},
  lrlab/.style={fill=white,inner sep=0.6pt,font=\footnotesize\itshape,text=neutral},
  ar/.style={-{Stealth[length=2.4mm,width=2.4mm]},line width=1pt,draw=neutral!55}
]
\node[bx=bad,text width=32mm]  (audio) at (-20mm,0)
 {\textbf{Questioned recording}\\[1pt] {\scriptsize genuine or spoofed}};
\node[bx=good,text width=32mm] (ref)   at (20mm,0)
 {\textbf{Genuine reference}\\[1pt] {\scriptsize same speaker, real}};
\node[bx=good,line width=1pt,text width=64mm] (exp) at (0,-12mm)
 {\textbf{2 examiners}: listening~~\pf\,\pf\, \\[1pt]
  {\scriptsize\itshape\color{neutral!65}\textbf{1 examiner}: casework + F0/acoustic + tool $\to$ LR}};
\node[bx=good,text width=32mm] (leg) at (-20mm,-32mm)
 {All YourTTS spoofs flagged \cmark\\ {\scriptsize\color{neutral!70}\itshape full-spoof LRs $\leq 0.97$}};
\node[bx=bad,text width=32mm]  (el)  at (20mm,-32mm)
 {Some ElevenLabs spoofs not flagged \xmark\\ {\scriptsize\color{neutral!70}\itshape LRs overlap genuine-item LRs}};
\node[bx=s1,text width=64mm] (c3) at (0,-46mm)
 {\textbf{Similar detection drop in both languages}\\[1pt]
  {\scriptsize\color{neutral!70}\itshape English and Czech}};
\draw[ar] (audio.south) -- (audio.south |- exp.north);
\draw[ar] (ref.south)   -- (ref.south   |- exp.north);
\draw[ar] (leg.north |- exp.south) -- node[lrlab]{LR} (leg.north);
\draw[ar] (el.north  |- exp.south) -- node[lrlab]{LR} (el.north);
\draw[ar] (leg.south) -- (leg.south |- c3.north);
\draw[ar] (el.south)  -- (el.south  |- c3.north);
\node[ctag] at (exp.north west) {C1};
\node[ctag] at (el.north west)  {C2};
\node[ctag] at (c3.north west)  {C3};
\end{tikzpicture}
\caption{Overview of the study and its three contributions: examiner evaluation (C1), casework mechanism (C2), and cross-language consistency (C3).
The key contrast is between older open-source synthesis and modern commercial voice cloning.}
\label{fig:abstract}
\end{figure}
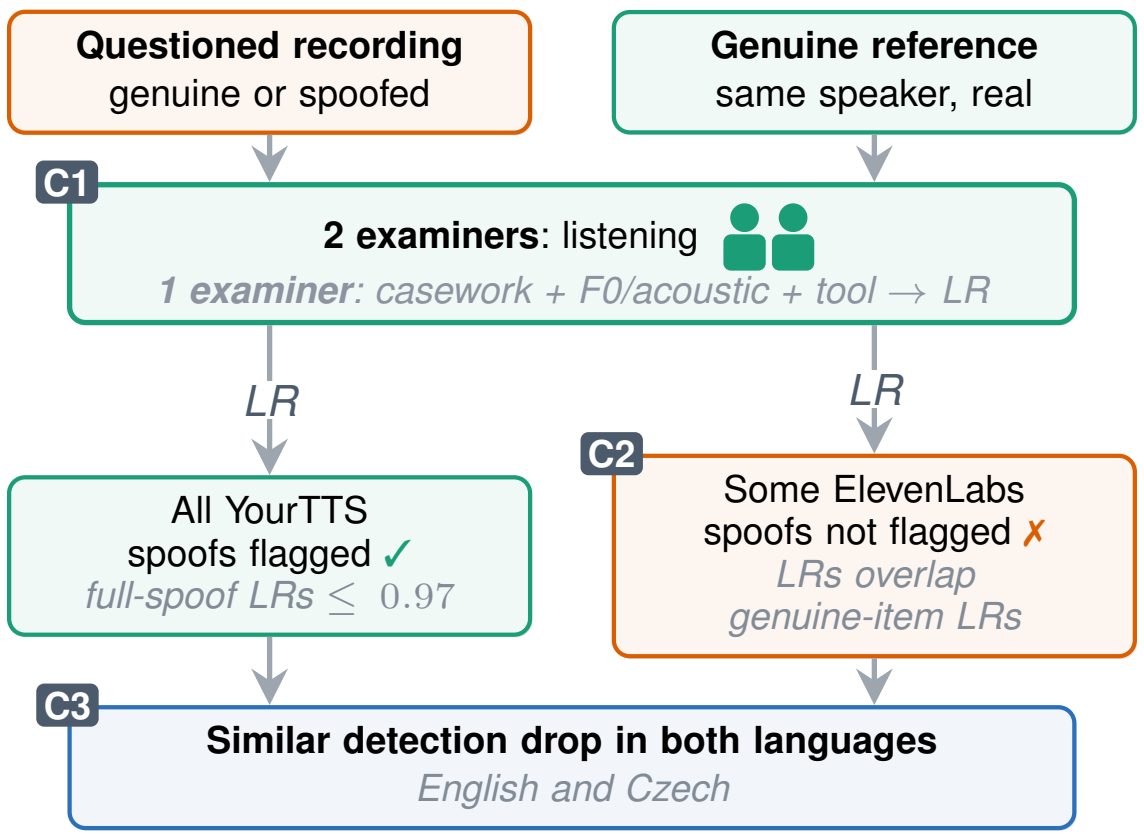
 
How forensic casework handles deepfake speech has rarely been measured, partly because practicing voice examiners are rare. We worked with two such examiners. Both completed the protocol-matched English listening task from a concurrent crowd study~\cite{trackingtrend} and a new Czech task. The Czech set compares genuine recordings with YourTTS (2022), an older open-source system~\cite{yourtts}, and ElevenLabs v1 (2024), a modern commercial service~\cite{elevenlabsdocs}. Examiner~A then processed the Czech recordings using the examiner's usual speaker-comparison tools and procedures, with a study-specific instruction to consider synthesis (Fig.~\ref{fig:abstract}). Because Czech is the examiners' casework language, the cross-language comparison tests whether second-language difficulty alone explains lower performance on ElevenLabs.

YourTTS was usually recognized as synthetic, whereas ElevenLabs was harder to detect in both languages. Some ElevenLabs recordings also received LRs within the range of the four genuine recordings and were not flagged for synthesis. Prior work has mostly measured software detectors or non-expert listening; we instead examine a casework-style process, including auditory and acoustic examination, with practicing examiners.
We make three contributions:

\noindent\textbf{(C1) Expert casework evidence.}
We connect sentence-level judgments from two practicing forensic voice examiners with the LR, local-anomaly notes, and final conclusion from Examiner~A's Czech workup. This exposes where synthetic-origin evidence can be lost between expert listening and the case-level conclusion.

\smallskip
\noindent\textbf{(C2) Casework failure mechanism.}
We show that modern synthetic speech can receive likelihood ratios that overlap those of the four genuine recordings because the LR assesses speaker identity, whereas the studied workflow provides no separate calibrated test for synthetic origin.

\smallskip
\noindent\textbf{(C3) Cross-language consistency.}
We observe a similar decline for ElevenLabs in English and Czech, suggesting that second-language difficulty is unlikely to be the sole explanation.

\section{Background and Related Work}\label{sec:related}
Relevant prior work spans forensic voice comparison, synthetic-speech detection, and human perception of full and partial spoofs.

\subsection{Forensic voice comparison and workflow}
Forensic voice comparison evaluates a questioned recording against a trusted reference from a known speaker. The examiner weighs the same-speaker and different-speaker propositions, usually through an LR. ENFSI guidelines specify this framework~\cite{enfsiguide}; Morrison and Enzinger survey the available comparison methods~\cite{morrison19,enfsi}; and Br\"{u}mmer and du~Preez formalize LR calibration and evaluation~\cite{brummer}.

In the casework we study, known in Czech and Central European practice as phonoscopy, the examiner combines three sources of evidence: auditory assessment of melody, tempo, rhythm, and voice quality; acoustic measurements such as F0 and spectral characteristics; and an automatic comparison score or LR. The cited voice-comparison methods and guidelines do not define a parallel procedure for synthetic-origin assessment. The participating examiner's routine process likewise includes no dedicated origin test: auditory and acoustic observations may suggest synthesis, but they produce no separately validated or calibrated origin result.
 
\subsection{Detection systems for forensic evidence}
Synthetic-speech detection asks the question that forensic voice comparison does not ask: whether the questioned recording is bona fide or synthetic.
Generic spoofing detectors, including systems developed for the ASVspoof challenges, show strong benchmark performance~\cite{asvspoof,asvspoof21,wang24_asvspoof}.
That performance is less reliable in deployment: detectors degrade on unseen domains, codecs, and attack conditions; benchmark design can inflate reported accuracy; and current commercial synthesis systems may be absent from training data~\cite{mullergen,berisha,datasetaudit,lastmile}.
These studies establish the need for separate synthesis screening, but not how to incorporate its output into a court-facing expert opinion.

Spoofing-aware speaker verification (SASV) and person-of-interest (POI) detection are closer to the forensic setting because they combine claimed-speaker matching with bona-fide or impersonation decisions~\cite{sasv,cuccovillo2026neuralencodingdetectionneed}.
Existing speaker-aware and POI methods use trusted reference material through target-speaker conditioning, speaker-profile comparison, phoneme-level analysis, retrieval-based difference modeling, or calibrated POI scores~\cite{osti1844063,liu23o_interspeech,Salvi2025poidetection,liu2026retrieval,roux2025calibratingpoi}.
However, these methods remain research systems evaluated under controlled protocols. They do not define how an examiner should combine identity and origin evidence or report separate, validated conclusions to a court. To our knowledge, no established forensic voice workflow currently does so. We therefore test the consequence of this gap in the studied casework workflow.
 
\subsection{Human perception and partial spoofing}
Human-listening studies have mostly tested non-expert participants rather than forensic voice examiners.
A meta-analysis of 56 detection studies ($N=86{,}155$) reports average deepfake-detection accuracy of 55.5 percent for non-experts~\cite{diel}, and controlled training experiments find no reliable improvement~\cite{mai}.
These results matter as a baseline, but they do not measure casework.
Partial spoofing creates a second problem.
Here, only one sentence or short segment is synthetic, and the rest of the recording is genuine.
ADD~2022~\cite{add22} and PartialSpoof~\cite{partialspoof} show that partial detection is a distinct problem.
For forensic comparison, the risk is that an utterance-level likelihood ratio can average over the entire recording, diluting a short synthetic segment with genuine surrounding speech.
Existing listening studies and partial-spoof benchmarks, therefore, leave the casework question open: what happens when a trained examiner, using an existing speaker-comparison workflow, evaluates such material?
 
\begin{figure*}[t!]
\centering
\resizebox{0.88\textwidth}{!}{
\sffamily
\definecolor{neutral}{HTML}{4C5B6B}
\definecolor{s1}{HTML}{2C6FBB}
\definecolor{s2}{HTML}{1B9E77}
\definecolor{s3}{HTML}{7570B3}

\def\W{6.2}\def\H{3.7}\def\HH{0.8}\def\GAP{0.55}

\newcommand{\icomic}{\tikz[baseline=0pt]{
  \draw[s1,line width=1.1pt,fill=s1!12,rounded corners=1.3mm] (-1.3mm,0.5mm) rectangle (1.3mm,4.8mm);
  \draw[s1,line width=1pt] (-2.5mm,1.7mm) arc (180:360:2.5mm);
  \draw[s1,line width=1pt] (0,-0.8mm) -- (0,-2.9mm);
  \draw[s1,line width=1pt] (-1.8mm,-2.9mm) -- (1.8mm,-2.9mm);}}
\newcommand{\icomag}{\tikz[baseline=0pt]{\draw[s2,line width=1.3pt] (0,1.8mm) circle (2.8mm);
  \draw[s2,line width=1.9pt,line cap=round] (2.0mm,-0.2mm) -- (4.9mm,-3.1mm);}}
\newcommand{\icogauge}{\tikz[baseline=0pt]{\draw[s3,line width=1.3pt] (0,0) arc (180:0:3.4mm);
  \draw[s3,line width=1.5pt,line cap=round] (3.4mm,0) -- (5.4mm,2.7mm); \fill[s3] (3.4mm,0) circle (0.6mm);}}

\newcommand{\panelhead}[3]{
  \draw[#1,line width=0.9pt,rounded corners=5pt,fill=#1!5] (0,0) rectangle (\W,-\H);
  \begin{scope}
    \clip[rounded corners=5pt] (0,0) rectangle (\W,-\H);
    \fill[#1] (0,0) rectangle (\W,-\HH);
  \end{scope}
  \node[circle,fill=white,text=#1,font=\bfseries\footnotesize,inner sep=0pt,minimum size=4.6mm]
    at (0.48,-0.41) {#2};
  \node[anchor=west,text=white,font=\bfseries\small] at (0.82,-0.41) {#3};
}
\begin{tikzpicture}[
  font=\footnotesize,
  bodyL/.style={anchor=north west,align=left,text=neutral,font=\footnotesize,text width=57mm,
                execute at begin node={\hyphenpenalty=10000\exhyphenpenalty=10000}},
  chip2/.style={rounded corners=3pt,draw=s2,fill=s2!8,line width=0.6pt,text=neutral,
                font=\scriptsize,align=left,text width=56mm,inner sep=3pt,anchor=north,
                execute at begin node={\hyphenpenalty=10000\exhyphenpenalty=10000}},
  ar/.style={-{Stealth[length=3.0mm,width=3.0mm]},line width=1.6pt,draw=neutral!55},
]

\begin{scope}[xshift=0cm]
  \panelhead{s1}{1}{Two stimulus sets}
  \node[bodyL] at (0.32,-1.22)
    {$\bullet$~\textbf{EN} (comparison): 9 speakers $\times$ 2 systems\\
     \hspace*{2.6mm}\textit{YourTTS\,/\,ElevenLabs v1}\\[2pt]
     $\bullet$~\textbf{CS} (new): 4 speakers $\times$ 2 systems\\
     \hspace*{2.6mm}\textit{YourTTS\,/\,ElevenLabs v1}\\[2pt]
     $\bullet$~Each clip = 4 sentences;\\
     \hspace*{2.6mm}\textbf{Full} = all synthetic, \textbf{Partial} = 1 of 4};
\end{scope}

\begin{scope}[xshift=6.75cm]
  \panelhead{s2}{2}{Two study stages}
  \node[anchor=north,text=neutral,font=\footnotesize\bfseries] at (3.05,-0.93) {2 forensic examiners};
  \node[chip2] at (3.05,-1.38)
    {{\color{s2}\textbf{1 \textbullet\ listening task}}: both examiners, EN+CS\\
     sentence labels + confidence};
  \node[chip2] at (3.05,-2.45)
    {{\color{s2}\textbf{2 \textbullet\ workup}}: Examiner A, CS, with tools\\
     genuine reference $\rightarrow$ \textbf{LR + conclusion}};
\end{scope}

\begin{scope}[xshift=13.5cm]
  \panelhead{s3}{3}{What we measure}
  \node[bodyL] at (0.32,-1.28)
    {$\bullet$~\textbf{Per recording:} LR, local notes, and\\ \hspace*{2.6mm}final conclusion\\[5pt]
     $\bullet$~\textbf{Per sentence:} was synthesis detected?\\[5pt]
     $\bullet$~\textbf{Wrong calls:} confident or unsure?};
\end{scope}

\draw[ar] (\W,-\H/2)             -- ++(\GAP,0);
\draw[ar] (2*\W+\GAP,-\H/2)      -- ++(\GAP,0);
\end{tikzpicture}}
\caption{Two-stage expert case-study design and measured outputs.}
\label{fig:method}
\end{figure*}
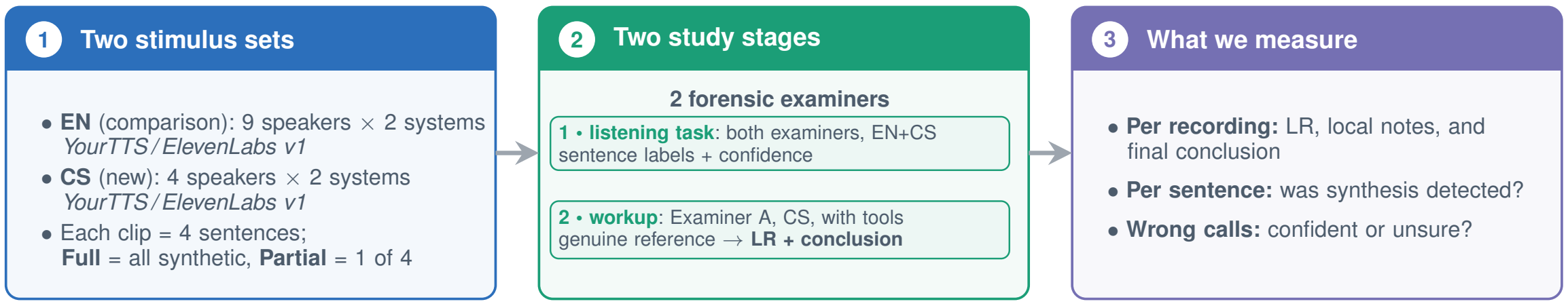

\section{Study Design and Evaluation Protocol}\label{sec:material}
We use an expert case-study design. Both examiners completed the English and Czech parts of a listening task. Examiner~A then performed a forensic workup of all twenty Czech recordings. Fig.~\ref{fig:method} summarizes this sequence.

\subsection{Test recordings.} 
Each recording contains four sentences. A genuine recording retains the original human speech; a full spoof synthesizes all four sentences, whereas a partial spoof replaces one complete genuine sentence with its synthetic counterpart.
The \emph{English set} is drawn from a concurrent companion listening study~\cite{trackingtrend} and contains nine English-speaking public figures. The \emph{Czech set} is newly constructed using the same recording structure and contains four Czech public figures. The Czech speakers formed a convenience sample selected based on the availability of sufficient public speech for cloning, test recordings, and a separate reference recording.
Both sets contain two synthesis conditions: YourTTS (2022)~\cite{yourtts} and ElevenLabs v1 (2024)~\cite{elevenlabsdocs}. We retained these systems from the companion study to preserve comparability and contrast older open-source synthesis with modern commercial voice cloning under the same protocol. The year labels indicate the release eras of the tool families rather than exact model checkpoints or recording-generation dates. For each speaker, the analyzed material therefore comprises five recordings: one genuine original, two full spoofs, and two partial spoofs. This yields 45 English and 20 Czech recordings.
Each synthesis system used approximately three minutes of target-speaker speech for zero-shot cloning, without speaker-specific fine-tuning. The ElevenLabs samples were generated in 2024 using the default settings of the v1-era commercial service. All recordings were delivered at 44.1 kHz and processed identically for volume normalization and background noise reduction.
 
\subsection{Examiners.}
Two practicing forensic voice examiners participated. At the time of the study, only three examiners in the Czech Republic routinely performed this type of forensic speaker-comparison casework; two participated in our study. Participation was uncompensated, and each examiner spent approximately two to three hours completing the English and Czech parts of the listening task. Both issue Czech-language speaker-comparison opinions for law-enforcement casework, making Czech their casework language and English a second-language comparison. Examiner~A contributed additional time for the Czech workup. This intensive access supports case-level analysis of a failure mechanism, not population-level estimates of examiner performance. Because the community is small and operationally sensitive, we report no employment details. We also withhold the name of the speaker-comparison software. The system compares x-vector embeddings~\cite{x-vector} and returns a likelihood ratio.

 \subsection{Tasks.} 
\subsubsection*{Task 1: online listening task}
The task was administered online, and headphone use was required. The examiners listened without forensic tools and without trusted speaker references. For each sentence, they classified the speech as genuine or synthetic and rated their confidence on a five-point scale: uncertain, low, medium, high, or full certainty. For summary counts, we group uncertain and low as low certainty, retain medium as medium certainty, and group high and full as high certainty.
The English part used the same recordings and protocol as the companion crowd study~\cite{trackingtrend}, allowing for contextual comparison with its reported aggregate results for non-expert listeners. We use only those aggregate results, not the individual participant responses. The companion study reports the non-expert listening experiment but does not include the forensic examiner results, the Czech casework workup, the likelihood ratio analysis, or the case-level conclusions reported in this work.
 
\subsubsection*{Task 2: forensic workup} Examiner~A additionally processed all twenty Czech recordings using the examiner's usual speaker-comparison tools and procedures. The examiner had completed the listening task but did not have access to the earlier responses or notes during the workup.
The examiner knew that the study concerned genuine, fully synthetic, and partially synthetic recordings and was instructed to consider possible synthetic origin in the overall casework conclusion. 
The examiner was not provided with the condition of each recording, the synthesis system, the manipulated sentence, or the item-level ground truth.
For each case, the test recording served as the questioned recording.
The trusted reference was a separate bona fide recording of the same speaker, taken from an additional reference video, not the genuine original item from the test set.
The examiner recorded an auditory impression, F0 and F0-variance measurements where available, the speaker-comparison likelihood ratio returned by the forensic tool, notes on any suspected local anomaly, and an overall conclusion.
The conclusion captured whether the examiner raised possible synthesis and was expressed in free-form casework language rather than as a forced choice. The speaker-comparison system computed each LR from the questioned recording and the separate bona fide reference, and the examiner interpreted the LR relative to 1. After item conditions were revealed, we compared the resulting LRs with the range observed for the four genuine test recordings. Auditory and acoustic observations could override the LR in the examiner's conclusion.

After the workup, a one-hour unrecorded debrief clarified the examiner's existing workup notes but yielded no additional data; we do not analyze it as evidence.
 
\subsection{Evaluation and Scope}

We analyze the listening task and the forensic workup separately. For the listening task, we pool both examiners and report class-specific sentence-classification rates. For spoof conditions, the rate is the proportion of synthetic sentences labeled synthetic; for genuine recordings, it is the proportion of genuine sentences labeled genuine. Using the same synthetic-sentence denominator makes full- and partial-spoof detection rates directly comparable. We report the corresponding counts with each rate. We retain condition-specific F1 scores only for direct within-condition comparisons with the companion study.

For the forensic workup, we report three distinct outputs: the whole-recording likelihood ratio, notes identifying possible local anomalies, and the final case-level conclusion. We binarize the free-form conclusions for analysis. Synthesis-leaning conclusions, including hedged formulations such as ``possible synthesis'' or ``synthesis or channel effect,'' are classified as \emph{flagged}; all other conclusions are classified as \emph{not flagged}. A recording that was not flagged was not thereby established as human-origin speech. Local anomaly notes are scored separately because an examiner may identify a suspicious segment without changing the recording's overall conclusion.

The participating examiner reported that routine speaker-comparison casework does not include a separate test of synthetic origin. For this study, however, the examiner was explicitly informed that the recordings could be genuine, fully synthetic, or partially synthetic and was instructed to consider synthesis when forming the final conclusion. The experiment, therefore, asks whether the existing speaker-comparison workflow can reveal synthesis when that possibility is made salient; it does not evaluate a dedicated synthesis-detection procedure or estimate performance in routine, unprompted casework.

Because the study contains two examiners and one workup, the reported scores and counts describe these cases; they do not estimate how often other examiners would make the same decisions. Sentence judgments repeat within speakers and examiners, so treating them as independent observations would inflate the sample size. The English and Czech sets use different speakers, so their comparison tests whether the pattern is consistent, not whether performance is statistically equivalent. Reference duration varied across speakers but was fixed within each speaker, and we did not inspect the speaker-comparison system's internal model or calibration. The study therefore establishes that the failure mechanism can occur, but not its prevalence in routine casework.

\subsection{Ethics and release.} The study was approved by the ethics committee at Brno University of Technology (decision EK:2024-002). The examiners gave informed consent. We collected no demographic data about the examiners and withheld their identities because the professional community is small. Synthetic recordings were created from public recordings of public figures for internal research. We do not release the recordings or prompts because the cloned voices could be misused.

\begin{table}[t]
\centering
\caption{Pooled sentence-classification rates for the two examiners, shown as rate (correct/total). For spoof conditions, the rate measures synthetic-sentence detection; for genuine recordings, it measures genuine-sentence classification.}
\label{tab:encs}
\begin{tabular}{lcc}
\toprule
Condition & English & Czech \\
\midrule

\multicolumn{3}{l}{\emph{Genuine-sentence classification}} \\
Genuine recordings & 0.61 (44/72) & 0.66 (21/32) \\

\addlinespace
\multicolumn{3}{l}{\emph{Synthetic-sentence detection: full spoof}} \\
YourTTS (2022)             & 0.94 (68/72) & 0.84 (27/32) \\
ElevenLabs (2024)          & 0.65 (47/72) & 0.63 (20/32) \\

\addlinespace
\multicolumn{3}{l}{\emph{Synthetic-sentence detection: partial spoof}} \\
YourTTS (2022)            & 0.94 (17/18) & 1.00 (8/8) \\
ElevenLabs (2024)          & 0.39 (7/18) & 0.63 (5/8) \\

\bottomrule
\end{tabular}
\end{table}

\section{Results}\label{sec:results}
We first report whether the examiners heard spoofing in English and Czech, then examine the casework outcomes for the Czech material.

\subsection{Listening task: cross-language and examiner outcomes}\label{sec:crosslang}
Both examiners completed the listening task on the companion study's English recordings and on the Czech set (Table~\ref{tab:encs}).
In both languages and both spoof types, the examiners detected a smaller proportion of ElevenLabs synthetic sentences than YourTTS synthetic sentences.
For full spoofs, detection decreased from 0.94 to 0.65 in English and from 0.84 to 0.63 in Czech, the examiners' casework language, when moving from YourTTS to ElevenLabs.
The decline in both languages is inconsistent with second-language difficulty as the sole explanation, even though the language sets include different speakers.

\subsubsection*{Czech examiner outcomes}
The pooled scores hide a marked split between the examiners (Table~\ref{tab:outcomes}). Examiner~A classified every genuine sentence correctly but missed most ElevenLabs full-spoof sentences. Examiner~B detected every such sentence but classified most genuine sentences as synthetic. Their partial-spoof errors showed the same contrast: Examiner~A missed three of four ElevenLabs synthetic sentences, whereas Examiner~B detected all synthetic sentences but misclassified 7/12 surrounding genuine sentences under each system.

Crucially, neither examiner made a high-certainty error. All nineteen errors by Examiner~A had low certainty. Of Examiner~B's twenty-seven errors, twenty-five had low certainty, and two had medium certainty; both occurred on YourTTS full spoofs. The error counts, therefore, reflect uncertain forced choices rather than firm false conclusions. With two examiners, these results do not establish a population-level trade-off. They instead show two possible failure modes: missing a modern clone and flagging genuine speech as synthetic.
 
\begin{table}[t]
\centering
\caption{Correct sentence-level classifications in the Czech listening task
(16 sentences per condition). Green marks perfect classification; red marks
fewer than half correct. The last row gives total errors, with medium- or
high-certainty errors in parentheses.}
\label{tab:outcomes}
\begin{tabular}{lcc}
\toprule
Condition & Examiner A & Examiner B \\
\midrule
Genuine originals
    & \cellcolor{goodbg}16/16
    & \cellcolor{flagbg}5/16 \\
YourTTS full spoof
    & 13/16
    & 14/16 \\
ElevenLabs full spoof
    & \cellcolor{flagbg}4/16
    & \cellcolor{goodbg}16/16 \\
YourTTS partial spoof
    & \cellcolor{goodbg}16/16
    & 9/16 \\
ElevenLabs partial spoof
    & 12/16
    & 9/16 \\
\midrule
Errors: total (medium/high certainty)
    & 19 (0)
    & 27 (2) \\
\bottomrule
\end{tabular}
\end{table}
 
\subsubsection*{Examiners and non-expert listeners}
Because both studies used the same English recordings and protocol, we additionally compare their condition-specific F1 scores within each condition; this is not a population comparison~\cite{trackingtrend}. For full spoofs, the examiners scored 0.97 on YourTTS and 0.79 on ElevenLabs, compared with 0.92 and 0.48 for the 82 non-expert listeners. For partial spoofs, the corresponding scores were 0.68 and 0.35 for the examiners and 0.90 and 0.29 for the non-experts. The examiners' clearest advantage was therefore limited to ElevenLabs full spoofs; they did not perform better across all conditions.

\subsection{Forensic workup: likelihood ratios and conclusions}

\begin{figure}[t]
\centering
\resizebox{\columnwidth}{!}{
\sffamily
\definecolor{neutral}{HTML}{4C5B6B}
\definecolor{band}{HTML}{1B7E70}
\definecolor{ptcol}{HTML}{34506B}
\begin{tikzpicture}[font=\scriptsize,yscale=0.58,
  pt/.style={circle,inner sep=0pt,minimum size=2.3mm,line width=0.8pt,draw=ptcol},
  hit/.style={pt,fill=ptcol},                    
  miss/.style={pt,fill=white},                   
  seg/.style={circle,inner sep=0pt,minimum size=3.3mm,draw=neutral,line width=0.6pt}]

\def\XA{0.9}\def\XB{2.5}\def\XC{4.1}\def\XD{5.7}\def\XE{7.3}   
\def\jA{-0.33}\def\jB{-0.11}\def\jC{0.11}\def\jD{0.33}         

\fill[band!10] (0.15,4.245) rectangle (8.05,5.295);
\node[anchor=south east,text=band!80!black,font=\tiny\itshape,inner sep=1pt] at (8.05,5.32) {observed genuine-item LR interval, (n=4)};

\draw[neutral,line width=0.6pt] (0.15,-0.15) -- (0.15,5.7);
\draw[neutral,line width=0.6pt] (0.15,-0.15) -- (8.05,-0.15);
\foreach \y/\l in {0/$10^{-8}$,1/$10^{-6}$,2/$10^{-4}$,3/$10^{-2}$,4/$1$,5/$10^{2}$}{
  \draw[neutral,line width=0.5pt] (0.08,\y) -- (0.15,\y);
  \node[anchor=east,text=neutral] at (0.05,\y) {\l};}
\node[rotate=90,text=neutral,anchor=south] at (-0.85,2.75) {Likelihood ratio, log scale};
\draw[neutral!60,dashed,line width=0.6pt] (0.15,4) -- (8.05,4);
\node[anchor=west,text=neutral!70,font=\tiny] at (8.05,4) {LR${=}1$};

\node[align=center,text=neutral,anchor=north] at (\XA,-0.3) {Genuine\\original};
\node[align=center,text=neutral,anchor=north] at (\XB,-0.3) {YourTTS\\full};
\node[align=center,text=neutral,anchor=north] at (\XC,-0.3) {YourTTS\\partial};
\node[align=center,text=neutral,anchor=north] at (\XD,-0.3) {ElevenLabs\\full};
\node[align=center,text=neutral,anchor=north] at (\XE,-0.3) {ElevenLabs\\partial};

\node[miss] at (\XA+\jA,4.98) {};
\node[miss] at (\XA+\jB,5.295) {};
\node[miss] at (\XA+\jC,4.85) {};
\node[miss] at (\XA+\jD,4.245) {};

\node[hit] at (\XB+\jA,3.81) {};
\node[hit] at (\XB+\jB,2.0) {};
\node[hit] at (\XB+\jC,0.47) {};
\node[hit] at (\XB+\jD,3.99) {};

\node[seg] at (\XC+\jA,5.17) {}; \node[hit] at (\XC+\jA,5.17) {}; 
\node[anchor=south,text=neutral,font=\tiny] at (\XC+\jA,5.30) {$\dagger$};
\node[seg] at (\XC+\jB,3.475) {}; \node[hit] at (\XC+\jB,3.475) {};
\node[seg] at (\XC+\jC,3.75)  {}; \node[hit] at (\XC+\jC,3.75) {};
\node[seg] at (\XC+\jD,4.21)  {}; \node[hit] at (\XC+\jD,4.21) {};

\node[miss] at (\XD+\jA,4.76) {};                      
\node[miss] at (\XD+\jB,4.59) {};                      
\node[hit]  at (\XD+\jC,2.68) {};                      
\node[hit]  at (\XD+\jD,4.33) {};                      

\node[miss] at (\XE+\jA,4.97) {};
\node[seg] at (\XE+\jB,4.67) {}; \node[hit] at (\XE+\jB,4.67) {};
\node[miss] at (\XE+\jC,4.73) {};
\node[miss] at (\XE+\jD,4.41) {};

\draw[neutral!40,rounded corners=2pt,fill=white] (4.2,0.32) rectangle (8.05,2.18);
\begin{scope}[font=\tiny]
  \node[hit,minimum size=1.9mm]  at (4.45,1.72) {};
  \node[anchor=west,text=neutral] at (4.63,1.72) {final conclusion: flagged};
  \node[miss,minimum size=1.9mm] at (4.45,1.14) {};
  \node[anchor=west,text=neutral] at (4.63,1.14) {final conclusion: not flagged};
  \node[seg,minimum size=3.1mm]  at (4.45,0.58) {};\node[miss,minimum size=1.7mm] at (4.45,0.58) {};
  \node[anchor=west,text=neutral] at (4.70,0.58) {local note: possible synthesis};
\end{scope}
\end{tikzpicture}}
\caption{Per-recording likelihood ratio, local synthesis notes, and binarized final conclusion from Examiner~A's forensic workup. LR$>$1 supports the same-speaker proposition. None of the four genuine recordings was flagged. Among full spoofs, 2/8 were not flagged, and 6/8 were; among partial spoofs, 3/8 were not flagged, and 5/8 were. All unflagged spoofs were generated with ElevenLabs.\\ $\dagger$:~flagged because of a local artifact despite a whole-recording LR of 215.}
\label{fig:lr}
\end{figure}
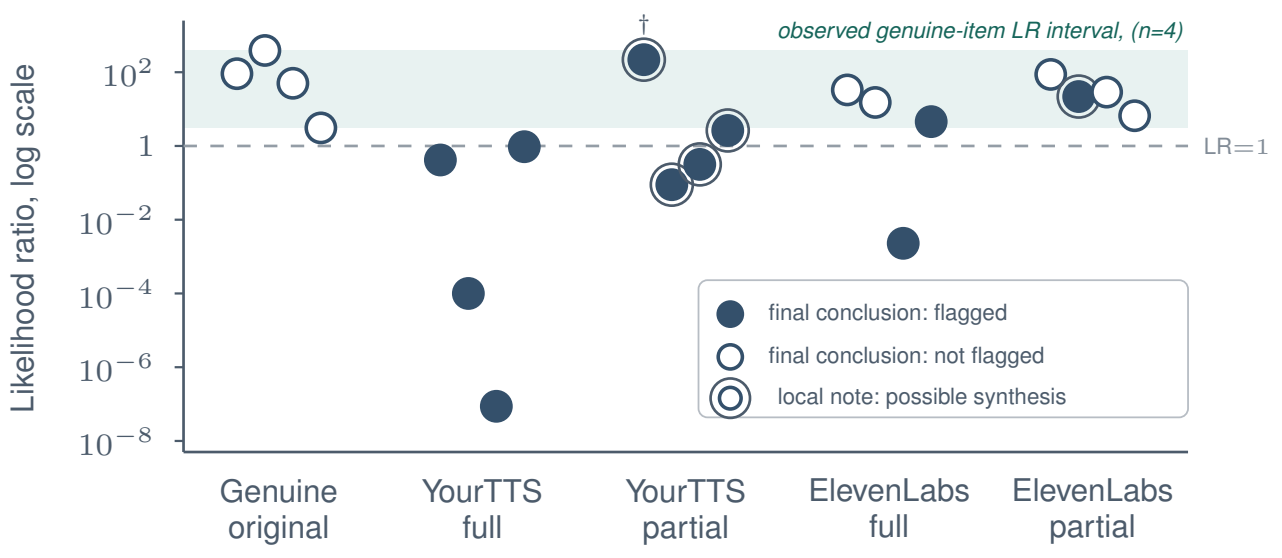

Fig.~\ref{fig:lr} reports the whole-recording LR, local anomaly notes, and final conclusion from the study workup.

For genuine recordings and YourTTS full spoofs, the likelihood ratios separate cleanly. The four genuine recordings receive LRs from 3.1 to 388. The four YourTTS full spoofs receive LRs from $9 \times 10^{-8}$ to 0.97, all below the neutral point LR = 1, and all four receive synthesis-leaning conclusions.

ElevenLabs full spoofs do not separate in the same way. For two of the four speakers, the LRs are 32.7 and 14.9, which overlap with the LRs of the four genuine recordings, and neither recording is flagged for synthesis in the final conclusion. A third recording receives weak same-speaker support (LR = 4.6) but is flagged, and the fourth receives an LR of 0.002 and is also flagged. Thus, two of the four ElevenLabs full spoofs are not flagged, compared with none of the four YourTTS full spoofs.

One ElevenLabs full spoof was not flagged for synthesis despite acoustically differing from its reference: the mean fundamental frequency increased from approximately 88 to 114~Hz. Its LR of 32.7 fell within the range observed for the four genuine recordings. The LR alone, therefore, did not reveal its synthetic origin.

Partial spoofs expose a second limitation: the whole-recording LR can remain high when only one sentence is synthetic. All four ElevenLabs partial spoofs receive LRs that overlap those of the four genuine recordings, ranging from 6.6 to 85.7, but one receives a synthesis-leaning final conclusion. The other three are not flagged. All four of the YourTTS partial spoofs are flagged.

Possible synthesis was noted locally in five of the eight partial spoofs. These five were flagged in the binarized final conclusion, while the remaining three were not. The outputs remain conceptually distinct: local notes mark suspicious regions, whereas the free-form conclusion determines the case-level score. In one YourTTS case, an artifact between 13 and 17 seconds overrode a whole-recording LR of 215.

The high partial-spoof LRs are consistent with dilution at the recording level. A partial spoof consists of one synthetic sentence surrounded by three genuine sentences from the target speaker, whereas the LR is computed over the entire recording. Listening may identify a suspicious region, but the studied workflow provides no local LR for evaluating that region separately.

\section{Discussion and Outlook}\label{sec:discussion}

Success on YourTTS did not carry over to ElevenLabs v1. Some ElevenLabs spoofs were not flagged for synthesis and received likelihood ratios within the range of the genuine recordings because the LR measures speaker identity, not synthetic origin. Local evidence sometimes overrode a high LR, as in the partial spoof with LR\,=\,215, but the workflow produced no separately validated result on origin. Speaker comparison alone, therefore, could not establish whether the speech was human.

The listening task exposed two further risks. Examiner~A classified all genuine sentences correctly but detected only 4/16 ElevenLabs full-spoof sentences. Examiner~B showed the opposite pattern, detecting all 16 ElevenLabs sentences but classifying only 5/16 genuine sentences correctly. None of their 46 errors were made with high certainty, and only two were made with medium certainty. With two examiners, this does not establish a general trade-off. It shows two possible outcomes under uncertainty: missing a modern clone and falsely flagging genuine speech.

Human listening is part of the existing forensic workflow evaluated here; we do not propose it as an effective countermeasure. The separate synthetic-origin assessment recommended below requires case-relevant validation and calibration, and its effectiveness was not evaluated in this study. The casework implication extends beyond adding a deepfake detector. Detector availability does not make a score suitable for forensic reporting: the score is informative only within an operating domain for which the system's discrimination has been validated and its numerical output has an established interpretation. Relevant dimensions include language, synthesis or attack family where it can be approximated, recording channel, codec, post-processing, duration, and background noise. Benchmark performance or a threshold transferred from another domain is insufficient because both discrimination and score interpretation can change under domain mismatch~\cite{mullergen,berisha,datasetaudit,lastmile,cuccovillo,roux2025calibratingpoi}.

This yields an actionable workflow. First, the laboratory should determine whether the questioned recording falls within the detector's validated domain. If it does, the laboratory should verify performance under the relevant conditions and calibrate any numerical score intended for interpretation. If it does not, the laboratory should construct independent, known-origin bona fide and synthetic materials that approximate the case conditions, then validate the detector and calibrate its score using representative data. Any fine-tuning requires separate held-out validation, and partial manipulations require segment-level as well as whole-recording testing~\cite{partialspoof,yangfsi}. If representative validation data cannot be obtained, origin should remain inconclusive rather than be inferred from a raw detector score. The speaker-identity LR and the origin conclusion should be reported separately.

This study shows that these failures can occur, not how often they occur. Future work should test additional examiners, speakers, languages, synthesis systems, and speaker-comparison tools, including workflows that incorporate an explicit synthesis check.

\section{Conclusion}\label{sec:conclusion}

Speaker comparison can support identity while leaving the recording origin unresolved. Our expert case study demonstrates this failure mechanism: modern synthetic recordings received same-speaker support and were not flagged for synthesis even when synthesis was explicitly considered. This complements machine- and human-detection studies by showing the downstream consequences of their limitations in casework. Deepfake detectors are not plug-in forensic tests. A numerical origin score is usable only after validation and calibration under case-relevant conditions. If those conditions are not met and representative validation data cannot be generated, the origin should remain inconclusive.

\section*{Acknowledgment}
The authors used ChatGPT (OpenAI) and Claude (Anthropic) for copy-editing and language refinement. All suggested changes were reviewed by the authors, who take full responsibility for the final text.

\bibliographystyle{IEEEtran}
\bibliography{references}

\end{document}